# pp Elastic Scattering at LHC
## Proton Structure   Outer Cloud – Inner Shell – Gluon Core

M. M. Islam and R. J. Luddy
University of Connecticut, Storrs, CT 06269 USA   islam@phys.uconn.edu   rjluddy@phys.uconn.edu

**Our investigation** of high energy pp and $\bar{p}p$ elastic scattering over the last twenty years has led us to the following physical picture of the proton (Fig.1)[1-3,7-8].

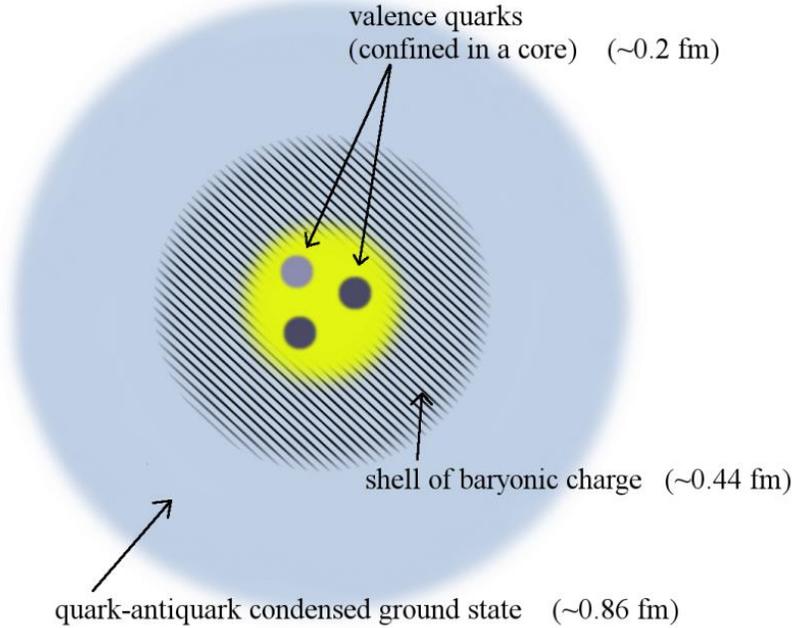

Fig.1. Physical picture of the proton.

The proton has three regions: an outer region consisting of a quark-antiquark ($q\bar{q}$) condensed ground state; an inner shell of baryonic charge; and a core of size 0.2 fm, where the valence quarks are confined.

The proton structure shown in Fig.1 leads to three main processes in elastic scattering as shown in Fig.2. The first process occurs in the small $|t|$ region, i.e., in the near forward direction, when the outer cloud of $q\bar{q}$ condensed ground state of one proton interacts with that of the other. It gives rise to diffraction scattering.

**Our Diffraction Amplitude** $T_D^+(s,t)$ in the impact parameter representation is taken to be

$$T_D^+(s,t) = i\, p\, W \int_0^\infty b\, db\, J_0(b\, q)\, \Gamma_D^+(s,b), \qquad (1)$$

and the profile function $\Gamma_D^+(s,b)$ is taken to be

$$\Gamma_D^+(s,b) = g(s)\left[\frac{1}{1+\frac{\exp(b-R)}{a}} + \frac{1}{1+\exp(-(b+R)/a)} - 1\right] = g(s)\frac{\sinh R/a}{\cosh R/a + \cosh b/a}; \qquad (2)$$

$g(s)$ is a function of s, $R = R(s) = R_0 + R_1(\ln s - i\frac{\pi}{2})$, and $a = a(s) = a_0 + a_1(\ln s - i\frac{\pi}{2})$. The parameters $R_0$, $R_1$, $a_0$ and $a_1$ are determined phenomenologically.

To obtain the coefficient $g(s)$ of the profile function $\Gamma_D^+(s,b)$ at $b = 0$, we notice

$$\Gamma_D^+(s,0) = g(s)\frac{\sinh R/a}{\cosh R/a + 1}. \qquad (3)$$

On the other hand, looking at the eikonal – we find   $\Gamma_D^+(s,0) = 1 - e^{i\chi_D^+(s,0)}. \qquad (4)$

Our phenomenological investigation leads us to the simple crossing-even term [2]

$$\Gamma_D^+(s,0) = 1 - \left(\eta_0 + \frac{c_0}{(s_+ e^{-i\pi/2})^\sigma}\right). \qquad (5)$$

Combining Eqs. (3) and (5) we obtain

$$g(s) = [1 - \eta_0 - \frac{c_0}{(s_+ e^{-i\pi/2})^\sigma}] \frac{\cosh R/a + 1}{\sinh R/a}. \tag{6}$$

At high energy $g(s) = [1 - \eta_0] \frac{\cosh R/a + 1}{\sinh R/a}$.

Furthermore, at high energy $\frac{R(s)}{a(s)} = \frac{R_0 + R_1 (\ln s - i\frac{\pi}{2})}{a_0 + a_1 (\ln s - i\frac{\pi}{2})} = \frac{R_0 + r\, a_0}{a_0 + a_1 (\ln s - i\frac{\pi}{2})} + r \simeq r$, where $r \equiv R_1/a_1$.

Then $g(s) = [1 - \eta_0] \frac{\cosh r + 1}{\sinh r}$ has become a constant – $\hat{g}$.

By changing the variable of integration in (1) from $b$ to $\zeta = b/a$ and rotating the line of integration of $\zeta$ to the real axis [1], we obtain

$$T_D^+(s,t) = i\, p\, W\, \hat{g}\, a^2(s) \int_0^\infty \zeta\, d\zeta\, J_0(\zeta\, q\, a) \frac{\sinh R/a}{\cosh R/a + \cosh \zeta}. \tag{7}$$

$$T_D^+(s,t) = i\, p\, W\, \hat{g}\, [a_0 + a_1 (\ln s - i\tfrac{\pi}{2})]^2 \int_0^\infty \zeta\, d\zeta\, J_0(\zeta\, q\, a) \frac{\sinh r}{\cosh r + \cosh \zeta}. \tag{8}$$

Defining $F(qa) \equiv \int_0^\infty \zeta\, d\zeta\, J_0(\zeta\, q\, a) \frac{\sinh r}{\cosh r + \cosh \zeta}$, we can write our diffraction amplitude as

$$T_D^+(s,t) = i\, p\, W\, \hat{g}\, [(a_0 + a_1 \ln s)^2 - i\, \pi\, a_1(a_0 + a_1 \ln s)]\, F(qa) \tag{9}$$

Our diffraction amplitude adheres to the well-known Asymptotic Properties:
(i) ✓ the observed increase of the pp total cross section with energy:
   $\sigma_{tot}(s) = 4\pi\, \hat{g}\, (a_0 + a_1 \ln s)^2\, F(0)$
   (recognized earlier as qualitative saturation of the Froissart-Martin Bound);
(ii) ✓ it leads to $\rho(s) = \pi\, a_1/(a_0 + a_1 \ln s)$;
(iii) ✓ $T_D^+(s,t) = i\, p\, W\, \hat{g}\, [(a_0 + a_1 \ln s)^2 - i\, \pi\, a_1(a_0 + a_1 \ln s)]\, F(qa)$
   (recognized earlier as Auberson-Kinoshita-Martin scaling);
(iv) ✓ pp and p̄p total cross sections are equal, and they have crossing-even scattering amplitudes.

**Beyond Diffraction Scattering** – the next dominant processes are multiple $\omega$-exchanges and valence quark-quark scattering via gluon-gluon interaction (Fig. 4). We also note that in pp and p̄p scattering – each $\omega$-exchange is accompanied by the glancing collision of the $q\bar{q}$ clouds of the protons (Fig. 3).

The combined amplitude due to $\omega$-exchanges and gluon-gluon interaction can be expressed in terms of a joint eikonal: $\chi_\omega(s,b) + \chi_{gg}(s,b)$. The amplitude $T_{\omega+gg}(s,t)$ is, however, screened by diffraction scattering. The resulting scattering amplitude is then

$$T_{\omega+gg}(s,t) = \left[\left(\eta_0 + \frac{c_0}{\left(s_+ e^{-i\frac{\pi}{2}}\right)^\sigma}\right) + i\left(\lambda_0 - \frac{d_0}{s^2}\right)\right] i\, p\, W \int_0^\infty b\, db\, J_0(b\, q)\, [1 - e^{i(\chi_\omega(s,b) + \chi_{gg}(s,b))}] \tag{10}$$

where the first two terms on the right-hand-side represent the screening effect of the clouds. Writing

$$[1 - e^{i(\chi_\omega(s,b) + \chi_{gg}(s,b))}] = [(1 - e^{i\chi_\omega(s,b)}) + e^{i\chi_\omega(s,b)}(1 - e^{i\chi_{gg}(s,b)})], \tag{11}$$

we approximate Eq. (11) as

$$T_{\omega+gg}(s,t) \simeq \left[\left(\eta_0 + \frac{c_0}{\left(s_+ e^{-i\frac{\pi}{2}}\right)^\sigma}\right) + i\left(\lambda_0 - \frac{d_0}{s^2}\right)\right] [T_\omega(s,t) + e^{i\chi_\omega(s,\hat{b})}\, T_{gg}(s,t)]. \tag{12}$$

$T_\omega(s,t)$ is the scattering amplitude due to multiple $\omega$-exchanges; $T_{gg}(s,t)$ is the gluon-gluon scattering amplitude (Fig. 4). $e^{i\chi_\omega(s,\hat{b})}$ is an average value that shows the additional screening of the scattering amplitude $T_{gg}(s,t)$, because of the baryonic-charge shell.

In similar fashion, for p̄p scattering, we find

$$\bar{T}_{\omega+gg}(s,t) \simeq \left[\left(\eta_0 + \frac{c_0}{(s_+ e^{-i\pi/2})^\sigma}\right) - i\left(\lambda_0 - \frac{d_0}{s_+^2}\right)\right] [\bar{T}_\omega(s,t) + e^{i\bar{\chi}_\omega(s,\hat{b})}\, \bar{T}_{gg}(s,t)] \tag{13}$$

We note that the explicit forms of both $T_{gg}(s,t)$ and $\bar{T}_{gg}(s,t)$ are the same and given by [8]:

$$T_{gg}(s,t) = \bar{T}_{gg}(s,t) = i\, s_+\, \gamma_{gg}(s_+\, e^{-i\pi/2})^\lambda \frac{\mathcal{F}^2(q_\perp, \lambda)}{\left(1 + \frac{q^2}{m^2}\right)^{2(\mu+1)}}. \tag{14}$$

**Polarization of the Clouds** – In our study of pp and p̄p scattering, we have realized—a new scattering amplitude occurs. To see this, let us go back to Fig. 1, which shows that the proton has a shell of baryonic charge and a core of valence quarks—also of baryonic charge. They are enclosed by the quark-antiquark cloud. The cloud becomes polarized, because its antiquarks are drawn toward the baryonic shell. In turn, a layer of polarization quarks appears near the boundary (Fig. 5). In pp near forward scattering, the two outer layers collide leading to a new scattering amplitude (positive). In p̄p near forward scattering, the outer polarization layer of the antiproton is of antiquarks and the polarization scattering amplitude is negative.

Thus, polarization of the clouds incorporates a small crossing-odd amplitude into our diffraction amplitude.

Our investigation has shown that the polarization amplitude $T_{pl}(s,t)$ can be expressed in terms of a profile function:

$$T_{pl}(s,t) = i\, p\, W \int_0^\infty b\, db\, J_0(b\, q)\, \Gamma_{Pl}^{+,-}(s,b), \qquad \text{pp (+), p̄p (-)} \tag{15}$$

and we find a suitable profile function to be

$$\Gamma_{Pl}^{+,-}(s,b) = \pm A\, e^{-b^2/B^2}\, J_0(b\, C) \qquad \text{pp (+), p̄p (-)} \tag{16}$$

with only three parameters: A, B and C.

### Determination of the Model Parameters  (The unit of energy used is 1.0 GeV.)

To describe diffraction, we need four parameters: $R_0 = 3.23$, $R_1 = 0.0508$, $a_0 = 0.375$ and $a_1 = 0.109$.
The values of the screening parameters are: $\eta_0 = 0.0594$, $\lambda_0 = 1.23$, $c_0 = 0.00$, $d_0 = 8.10$, $\sigma = 1.00$.
Four parameters involved the hard $\omega$-exchange amplitude: $\beta = 3.743$, $m_\omega = 0.655$, $\hat{\gamma} = 1.44$, $\hat{\theta} = 1.39$.
The screening terms $e^{i\,\chi_\omega(s,\hat{b})}$ and $e^{i\,\bar{\chi}_\omega(s,\hat{b})}$ have one additional parameter: $\hat{b}$. We find $\hat{b} = 0.00$.
The parameters found for the amplitude $T_{gg}(s,t)$ are $\gamma_{gg} = 0.019$, $\lambda = 0.221$, $\mu = \frac{1}{4}$, $m = 1.67$.
For the polarization amplitudes $T_{pl}^{+,-}(s,t)$, we find the parameters $A = 0.0236$, $B = 0.0034$, $C = 0.635$.

### Analysis

Our prediction for pp elastic $d\sigma/dt$ at 13 $TeV$ is shown in Fig. 6. The measurement has been done at the LHC by the TOTEM Collaboration. The TOTEM data shown is preliminary, and we have predicted the normalization of the TOTEM data, based on our model.

The most striking feature of the preliminary 13 $TeV$ TOTEM data is that there are no oscillations in $d\sigma/dt$ beyond the initial dip-bump structure. It shows a smooth falloff for large $|t|$, exactly as predicted by our model.

Our three-region model has always predicted a change of slope in $d\sigma/dt$ at a large value of $|t|$, where the $gg$-interactions begin to dominate the $\omega$-interactions. We may already be seeing this in the TOTEM 13 $TeV$ data above 3.1 $GeV^2$, and we see it in our calculation. However, the apparent leveling-off of the TOTEM data at large $|t|$ may be due to the lower statistics in that region.

Fig. 6 also shows our calculated pp $d\sigma/dt$ at 7 TeV along with the TOTEM data, as well as our calculated p̄p $d\sigma/dt$ at 1.96 TeV presented with the D0 data and the earlier Tevatron p̄p 1.80 TeV data.

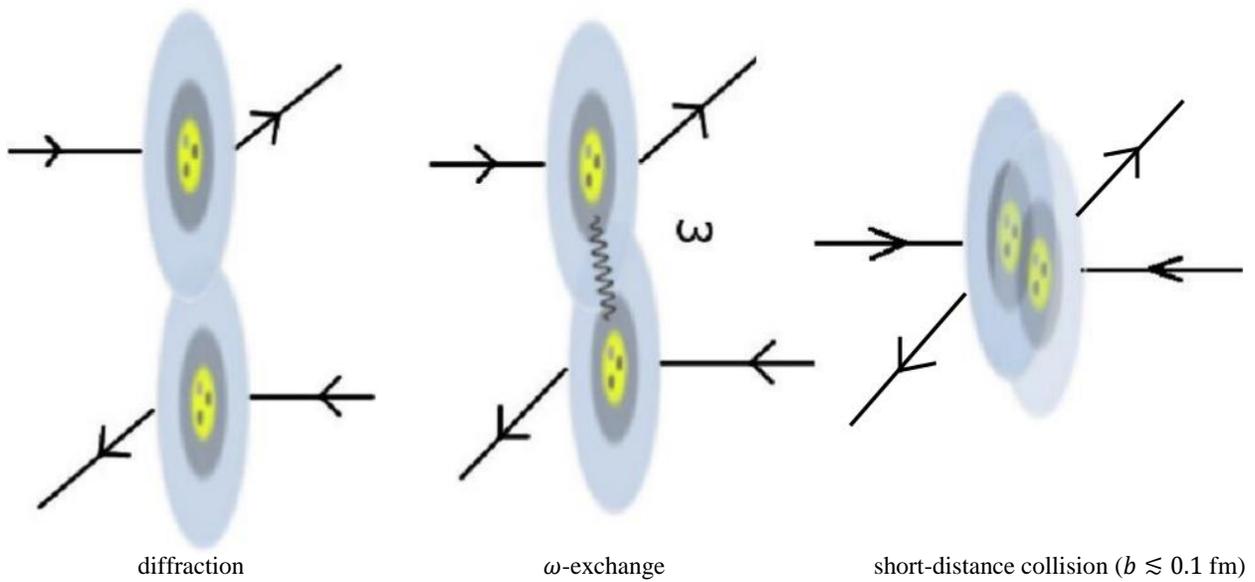

diffraction   ω-exchange   short-distance collision ($b \lesssim 0.1$ fm)

Fig. 2. Three elastic scattering processes.

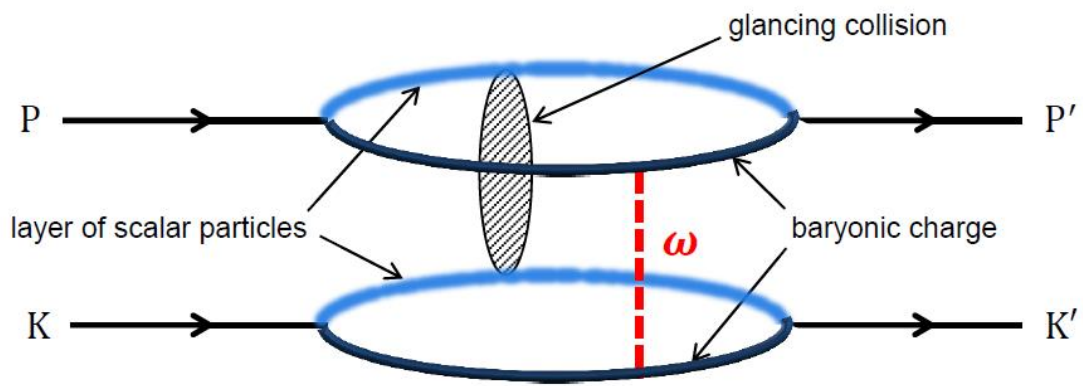

Fig.3. Single $\omega$-exchange accompanied by a glancing collision of scalar particles of one proton with those of the other proton (in momentum space).

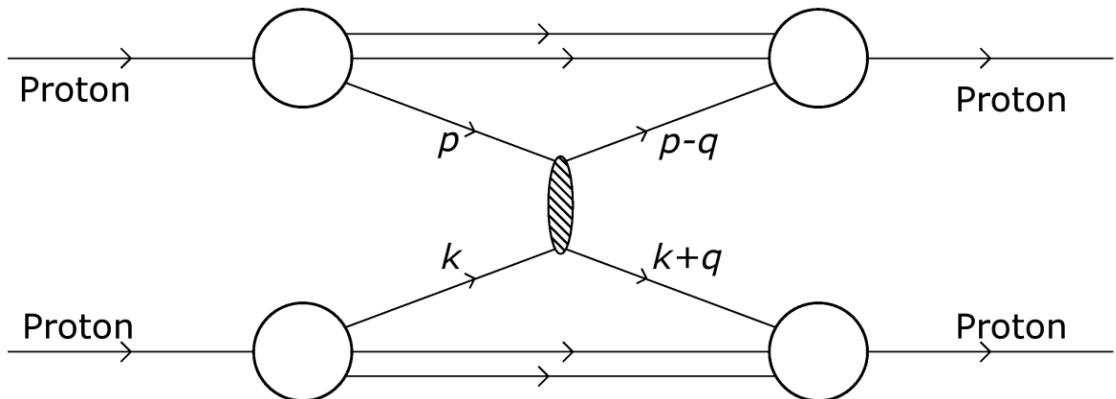

Fig.4. Hard collision of a valence quark from one proton with one from the other proton.

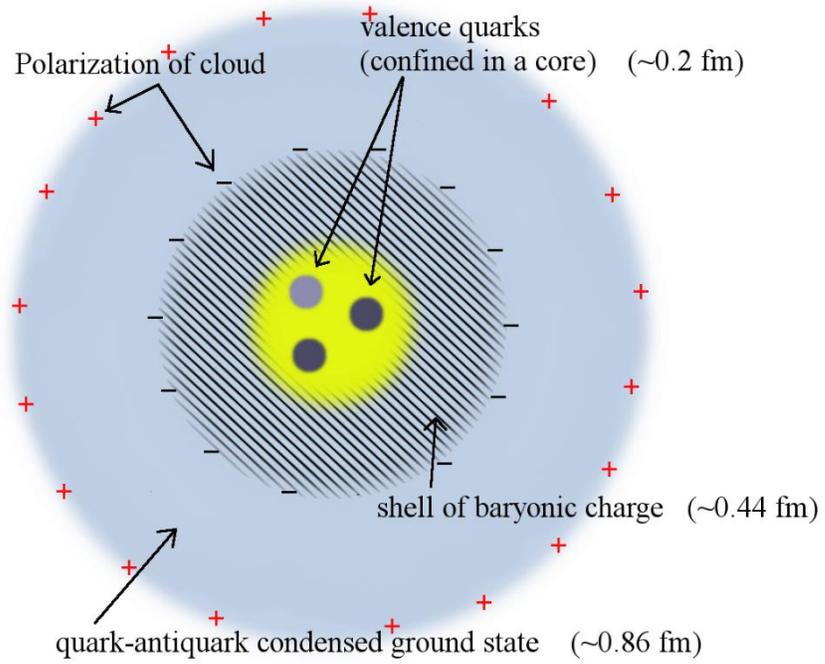

Fig.5. Physical picture of the proton – with polarization of the cloud shown.

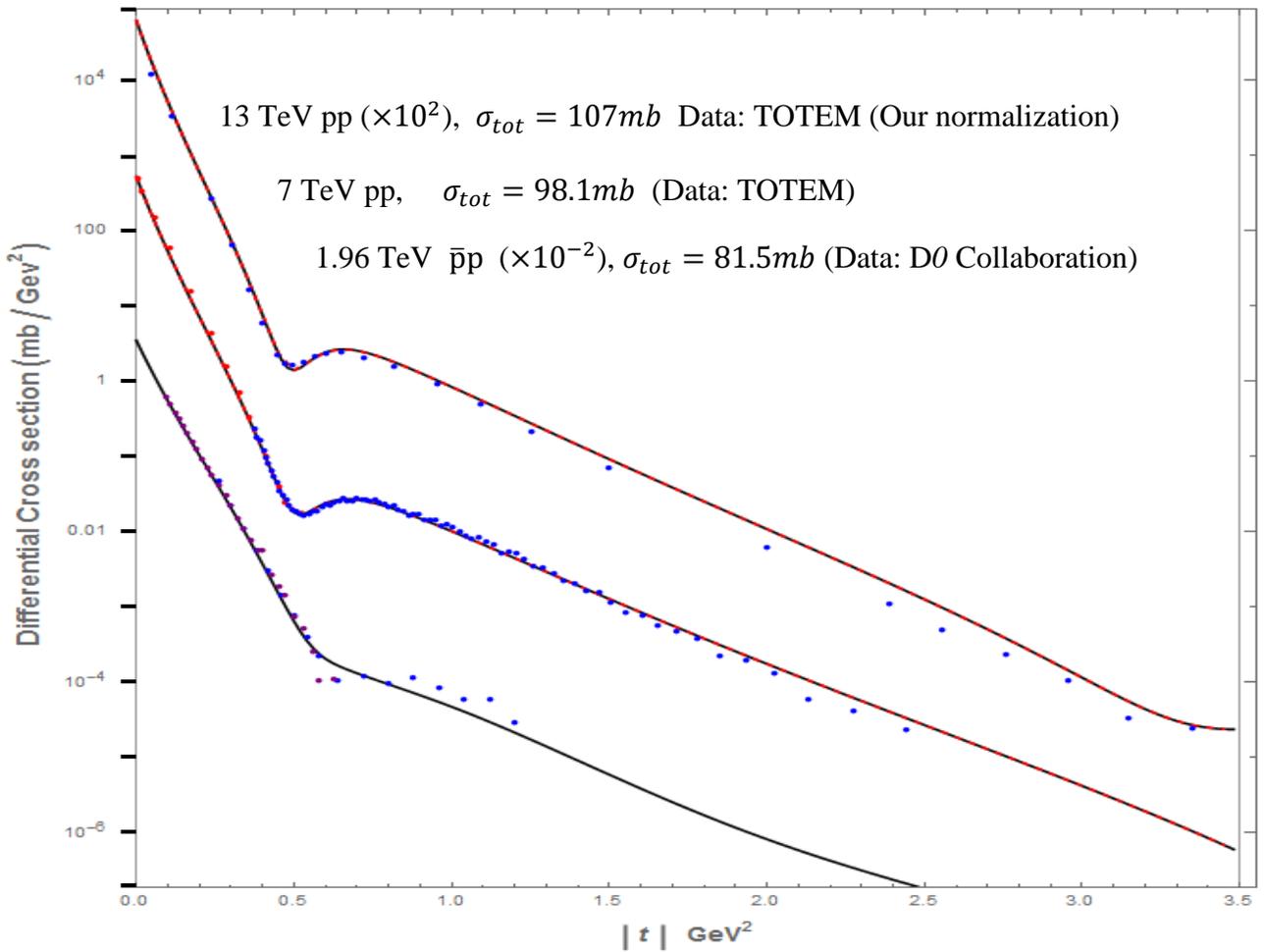

Fig.6. Our $d\sigma/dt$ prediction at $\sqrt{s}$ = 13 TeV – with preliminary data by the TOTEM Collab.[9]
Comparison of our $d\sigma/dt$ calculation at $\sqrt{s}$ = 7 TeV with the TOTEM Collab. measurements at LHC [10].
Comparison of our $d\sigma/dt$ calculation at $\sqrt{s}$ = 1.96 TeV with the D0 Collab. and 1.8 TeV data [11,12].